\documentstyle[12pt,eqsecnum,aps,fleqn,prb]{revtex}

\begin{document}

\title{The one-loop effective action and trace anomaly in four
dimensions}
\vspace{10mm}
\author{\\
  A. O. Barvinsky\hbox{$^{1,\rm a)}$},
  Yu. V. Gusev\hbox{$^{1,\rm b)}$},
  G. A. Vilkovisky\hbox{$^{2}$} \\and\\
  V. V. Zhytnikov\hbox{$^{1,\rm c)}$}
 }

\maketitle
%\vspace{5mm}
\begin{center}
$^{1}$ {Nuclear Safety Institute, Bolshaya Tulskaya 52, Moscow
113191,
Russia}\\$^{2}${Lebedev Physics Institute, Leninsky Prospect 53,
Moscow 117924,
Russia}
\end{center}
%\bigskip
%\bigskip
\vspace{10mm}

\begin{abstract}
The one-loop effective action for a generic set of quantum fields is
calculared
as a nonlocal expansion in powers of the curvatures (field
strengths). This
expansion is obtained to third order in the curvature. It is stressed
that the
covariant vertices are finite. The trace anomaly in four dimensions
is obtained
directly by varying the effective action. The nonlocal terms in the
action,
producing the anomaly, contain non-trivial functions of three
operator
arguments. The trace anomaly is derived also by making the conformal
transformation in the heat kernel.
\end{abstract}

\vspace{10mm}

\thispagestyle{empty}
\footnoterule
\begin{itemize}
\item[$^{\rm a)}$]On leave at the Department
of Physics, University of Alberta, Edmonton T6G 2J1, Canada
\item[$^{\rm b)}$]
On leave at the Department of Physics,
University of Manitoba,
Winnipeg R3T 2N2, Canada
\item[$^{\rm c)}$]
On leave at the Department of Physics of
National Central University,
Chung-li, Taiwan 320
\end{itemize}

\pagebreak

\section{Introduction}
\indent
The trace anomaly in the vacuum energy-momentum tensor of conformal
invariant
fields discovered long ago \cite{DDI,Duff,Brown,FV-conf} has
important
applications \cite{Pol,appl,FrolV,2dBH} and arouses presently a new
interest
\cite{Tomb,II,Mott,GAV-Strasb,DesSch,Osborn,JMP2}. The recent studies
are
motivated by the wish to extend two-dimensional results \cite{Mott}
and by the
fact that the anomaly in four dimensions may be associated with
important
physical effects \cite{Chris,FrolV,Page,II,GAV-Strasb}.

The most interesting feature of the trace anomaly is that it contains
an
information about the nonlocal structure of the effective action. It
is the
nonlocal effective action that gives rise to the physical effects
\cite{FrolV,Pol,Gospel,OstrV,2dBH,GAV-Strasb,CQG,Armen,CQGas}.
Unlike in two
dimensions \cite{Chris,Pol,FrolV,Gospel}, in higher dimensions the
trace
anomaly does not control the effective action completely. Two actions
producing
one and the same anomaly differ generally by an arbitrary conformal
invariant
functional. Furthermore, in higher dimensions the off-shell effective
action
can be (regularized and) calculated so that there will be no anomaly
\cite{Englert,FV-conf}. The difference between the anomalous and
non-anomalous
actions may be of importance for field's expectation values. No
physical test has
been proposed thus far to make a choice between
the two. Nevertheless, the trace anomaly may help to build a model
effective action
that would contain the essential nonlocal effects and could be used
for studying the
field's expectation values in four and higher dimensions.

Although there have been studies aimed at building nonlocal geometric
invariants associated with the anomaly
\cite{Riegert,Tomb,Mott,DesSch}, the
latter has never been derived from the effective action directly. The
purpose
of the present work is to fill this gap. As will be seen, the
nonlocal
structures that really appear in the effective action and give rise
to the
anomaly are more complicated than the ones used in the above
mentioned
constructions.

The one-loop effective action for a generic set of quantum fields has
recently
been calculated to third order in the curvature \cite{IV}.  The
method for this
calculation, the covariant perturbation theory, was developed in
Refs.\onlinecite{I,II,III}. In this method, the effective action is
obtained as
a nonlocal expansion in powers of a universal set of field strengths
(curvatures) characterizing a generic field model. To reproduce the
trace
anomaly in four dimensions, one needs this expansion to third order.
Here we
shall use the results of the report \onlinecite{IV} to display the
mechanism by
which the local trace anomaly emerges from the nonlocal effective
action.

The plan of the paper is as follows. In Sect.II we present the form
of the
nonlocal expansion for the effective action to third order in the
curvature,
and comment on its derivation by covariant perturbation theory. We
discuss also
the mechanism by which the finite covariant vertices are obtained.
Sect.III
contains a direct derivation of the trace anomaly from the nonlocal
effective
action. We show, in particular, that quadratic (in the curvature)
terms of the
action do not generate a local anomaly. Their contribution to the
anomaly, even
to the lowest nontrivial order, contains a certain nonlocal form
factor which
is cancelled only after adding the contribution of the covariant
vertices (cf.
Ref.\onlinecite{DesSch}). In Sect.IV, the trace anomaly is derived by
making
the conformal transformation of the heat kernel.  In this derivation,
we use
the results of calculation of the heat kernel by covariant
perturbation theory
\cite{II,IV,JMP2}. We show that the mechanism of cancellation of
nonlocal terms
leading to the local anomaly is the same as in two dimensions
\cite{JMP2}.
Namely, after the conformal transformation the heat kernel becomes a
total
derivative, and the anomaly is determined only by the early-time
asymptotic
behaviour.

\section{The effective action in covariant perturbation theory}
\indent
Here we consider the effective action in noncompact asymptotically
flat and
empty spacetime with the euclidean (positive-signature) metric, which
is
sufficient for finding the expectation-value equations of lorentzian
field
theory with the {\em in}-vacuum quantum state \cite{FrolV,II}. For a
generic
quantum field model it can be obtained by calculating loops with the
inverse
propagator -- the second-order operator
	\begin{eqnarray}
	H=g^{\mu\nu}\nabla_{\mu}\nabla_{\nu}+
	\Big(\hat P-\frac16\,R\hat 1\,\Big),
\label{2.1}
	\end{eqnarray}
acting on an arbitrary set of fields $\varphi^A(x)$. Here $A$ stands
for any
set of discrete
indices, and the hat indicates that the quantity is a matrix in the
corresponding vector space of field components:
	$\hat{1}=\delta^A{}_B,\,
	\hat{P}=P^A{}_B,\ {\rm etc.}$
Below the matrix trace operation will be denoted by ${\rm tr}$ :
	${\rm tr}\hat{1}=\delta^A{}_A$,
	${\rm tr}\hat{P}=P^A{}_A,\ {\rm etc.}$
In (1.2), $g_{\mu\nu}$ is a positive-definite metric
characterized by its Riemann curvature
$R^{\alpha\beta\mu\nu}$, for which we use the sign convention
$R^\mu_{\,\cdot\, \alpha\nu\beta}=\partial_\nu
\Gamma^\mu_{\alpha\beta}-\cdots,\  R_{\alpha\beta}=
R^\mu_{\,\cdot\, \alpha\mu\beta},\
R=g^{\alpha\beta}R_{\alpha\beta}$,
$\nabla_\mu$ is a covariant derivative (with respect to an
arbitrary connection) characterized by its commutator
curvature
	\begin{equation}
	(\nabla_\mu\nabla_\nu-\nabla_\nu\nabla_\mu)
	\varphi^A =
	{\cal R}^A{}_{B\mu\nu}\varphi^B,\hspace{7mm}
	{\cal R}^A{}_{B\mu\nu}\equiv\hat{\cal R}_{\mu\nu},
\label{2.3}
	\end{equation}
and $\hat{P}$ is an arbitrary matrix. The redefinition of the
potential in (\ref{2.1}) by inclusion of the term in the Ricci
scalar $R$ is a matter of convenience.

There are three independent inputs in the operator
(\ref{2.1}): $g^{\mu\nu}$, $\nabla_{\mu}$ and $\hat P$
- the metric contracting the second
derivatives, the connection which defines the covariant derivative
and the potential matrix. They may be regarded as background fields
to which correspond their field strengths or curvatures. There is the
Riemann curvature associated with $g_{\mu\nu}$, the commutator
curvature (\ref{2.3}) associated with $\nabla_{\mu}$ and the
potential
$\hat P$ which is its own "curvature". An important feature of the
asymptotically flat spacetime is that its purely gravitational
strength
boils down to the Ricci
tensor, because the Riemann tensor can always be eliminated via the
differentiated Bianchi identity \cite{II,JMP1} by iteratively
solving it for $R^{\alpha\beta\mu\nu}$ in terms of $R^{\mu\nu}$. This
iterational solution is uniquely determined  by the Green function
$1/\Box$
with
zero boundary conditions at infinity and starts with
	\begin{eqnarray}&& R^{\alpha\beta\mu\nu}=
	\frac12\,\frac1{\Box}\Big(
	 \nabla^\mu \nabla^\alpha R^{\nu\beta}
	+\nabla^\alpha \nabla^\mu R^{\nu\beta}
	-\nabla^\nu \nabla^\alpha R^{\mu\beta}
	-\nabla^\alpha \nabla^\nu R^{\mu\beta}
	\nonumber\\
	&&
	\qquad\quad
	-\nabla^\mu \nabla^\beta R^{\nu\alpha}
	-\nabla^\beta \nabla^\mu R^{\nu\alpha}
	+\nabla^\nu \nabla^\beta R^{\mu\alpha}
	+\nabla^\beta \nabla^\nu R^{\mu\alpha}
	\Big)
	+{\rm O}[R^2_{..}].
\label{2.3a}
	\end{eqnarray}
Thus, the full set of field strengths characterizing the operator
(\ref{2.1}), for which we shall use the collective notation $\Re$,
includes
	\begin{equation}
	\Re=(R_{\mu\nu},\;\hat{\cal R}_
	{\mu\nu},\;\hat P).                            \label{2.5}
	\end{equation}

At one-loop order the effective action is given by the expression
	\begin{eqnarray}
	-W=-\frac 12\,{\rm Tr \,ln}\,H+
	\int d^4x\,\delta^4(x,x)\,(...),
\label{2.4}
	\end{eqnarray}
where ${\rm Tr}$ as distinct from ${\rm tr}$ denotes the functional
trace
and the term with ellipses $(\dots)$ stands for the
contribution of the local functional measure,
proportional to the delta-function at  coincident
points. As shown in Ref.\onlinecite{Bern}, this contribution always
cancels the volume divergences of the loop
which otherwise would appear in (\ref{2.4}) in the form of
a divergent cosmological term. {\em For a massless
operator} (\ref{2.1}), the result of this cancellation is a
subtraction of the
term of zeroth
order in the curvature. The masslessness of the operator (\ref{2.1})
means
that, like the
Riemann and commutator curvatures, the potential $\hat{P}$
falls off at infinity of the manifold. For the
precise conditions of this fall off see Ref.\onlinecite{II}.

The effective action (\ref{2.4}) is an invariant functional of
background
fields. In the four-dimensional euclidean asymptotically flat
spacetime, it is
analytic in the curvatures (\ref{2.5}) \cite{JMP2}. It is, therefore,
expandable in the basis of nonlocal invariants
of $N$th order in  $\Re$ built in Ref.\onlinecite{JMP1} for $N=1,\,2$
and $3$. In covariant perturbation theory, the expansion goes in
powers of
these
curvatures but its coefficients ( the nonlocal form factors) are also
field dependent. The meaning of such an expansion is that the
effective action
is obtained with accuracy $O\,[\,\Re^N\,]$, i.e. {\it up to} $N$-th
power in
$\Re$.
Each term of the expansion contains $N$ curvatures $\Re$ {\em
explicitly} and
is
defined up to $O\,[\,\Re^{N+1}\,]$.

The result for the one-loop effective action (\ref{2.4}) to third
order in the
curvature
is of the form \cite{IV}
	\begin{eqnarray}
	-W&=&{\frac1{2(4\pi)^2}\int\! dx\, g^{1/2}\, \,{\rm tr}\,}
	\left\{\sum^{5}_{i=1}\gamma_i(-\Box_2)\Re_1\Re_2({i})\right.
	\nonumber\\
	&&\qquad\qquad\left.+\sum^{29}_{i=1}
	\Gamma_{i}(-\Box_1,-\Box_2,-\Box_3)
	\Re_1\Re_2\Re_3({i})+{\rm O}[\Re^4]\right\}.    \label{2.6}
	\end{eqnarray}
Here terms of zeroth and first order in the curvature
are omitted \cite{f0}. The quadratic
$\Re_1\Re_2({i}),\;i=1\;{\rm to}\;5,$ and cubic
$\Re_1\Re_2\Re_3({i}),\;i=1\;{\rm to}\;29,$ curvature invariants
here, as well as the conventions concerning the action of covariant
D'Alambertian
arguments $(\Box_1,\Box_2,\Box_3)$ on the curvatures
$(\Re_1,\Re_2,\Re_3)$ labelled by the corresponding numbers,
are presented and discussed in much detail in
Refs.\onlinecite{II,JMP1}.
Terms of second order in the curvature are given
by five quadratic structures
	\begin{eqnarray}
	\Re_1\Re_2({1})&=&
	R_{1\,\mu\nu} R_2^{\mu\nu}\hat{1},\nonumber\\
	\Re_1\Re_2({2})&=&R_1 R_2\hat{1},\nonumber\\
	\Re_1\Re_2({3})&=&\hat{P}_1 R_2,\nonumber\\
	\Re_1\Re_2({4})&=&\hat{P}_1\hat{P}_2,\nonumber\\
	\Re_1\Re_2({5})&=&\hat{\cal R}_{1\mu\nu}
	\hat{\cal R}_2^{\mu\nu}.
\label{2.7}
	\end{eqnarray}
and terms of third order by twenty nine cubic structures
$\Re_1\Re_2\Re_3({i})$, $i=1,...29$. Eleven of them contain no
derivatives
	\begin{eqnarray}
	\Re_1\Re_2\Re_3({1})&=&\hat{P}_1\hat{P}_2\hat{P}_3,
	\nonumber\\
	\Re_1\Re_2\Re_3({2})&=&\hat{\cal R}^{\ \mu}_{1\ \alpha}
	\hat{\cal R}^{\ \alpha}_{2\ \beta}
	\hat{\cal R}^{\ \beta}_{3\ \mu},\nonumber\\
	\Re_1\Re_2\Re_3({3})&=&\hat{\cal R}^{\mu\nu}_1
	\hat{\cal R}_{2\,\mu\nu}\hat{P}_3,\nonumber\\
	\Re_1\Re_2\Re_3({4})&=&R_1 R_2 \hat{P}_3,\nonumber\\
	\Re_1\Re_2\Re_3({5})&=&R_1^{\mu\nu}
	R_{2\,\mu\nu}\hat{P}_3,\nonumber\\
	\Re_1\Re_2\Re_3({6})&=&\hat{P}_1\hat{P}_2 R_3,
	\nonumber\\
	\Re_1\Re_2\Re_3({7})&=&R_1\hat{\cal R}^{\mu\nu}_2
	\hat{\cal R}_{3\,\mu\nu},\nonumber\\
	\Re_1\Re_2\Re_3({8})&=&R_1^{\alpha\beta}
	\hat{\cal R}_{2\,\alpha}^{\ \ \ \,\mu}
	\hat{\cal R}_{3\,\beta\mu},\nonumber\\
	\Re_1\Re_2\Re_3({9})&=&R_1 R_2 R_3\hat{1},
	\nonumber\\
	\Re_1\Re_2\Re_3({10})&=&R_{1\,\alpha}^\mu
	R_{2\,\beta}^{\alpha} R_{3\,\mu}^\beta\hat{1},
	\nonumber\\
	\Re_1\Re_2\Re_3({11})&=&
	R_1^{\mu\nu}R_{2\,\mu\nu}R_3\hat{1},
\label{2.8}
	\end{eqnarray}
fourteen structures contain two derivatives
	\begin{eqnarray}
	\Re_1\Re_2\Re_3({12})&=&
	\hat{\cal R}_1^{\alpha\beta}\nabla^\mu
	\hat{\cal R}_{2\mu\alpha}\nabla^\nu
	\hat{\cal R}_{3\nu\beta},\nonumber\\
	\Re_1\Re_2\Re_3({13})&=&\hat{\cal R}_1^{\mu\nu}
	\nabla_\mu\hat{P}_2\nabla_\nu\hat{P}_3,\nonumber\\
	\Re_1\Re_2\Re_3({14})&=&\nabla_\mu
	\hat{\cal R}_1^{\mu\alpha}\nabla^\nu
	\hat{\cal R}_{2\,\nu\alpha}\hat{P}_3,\nonumber\\
	\Re_1\Re_2\Re_3({15})&=&
	R_1^{\mu\nu}\nabla_\mu R_2\nabla_\nu
	\hat{P}_3,\nonumber\\
	\Re_1\Re_2\Re_3({16})&=&
	\nabla^\mu R_1^{\nu\alpha}\nabla_\nu
	R_{2\,\mu\alpha}\hat{P}_3,\nonumber\\
	\Re_1\Re_2\Re_3({17})&=&
	R_1^{\mu\nu}\nabla_\mu\nabla_\nu
	\hat{P}_2\hat{P}_3,\nonumber\\
	\Re_1\Re_2\Re_3({18})&=&R_{1\,\alpha\beta}\nabla_\mu
	\hat{\cal R}_2^{\mu\alpha}\nabla_\nu
	\hat{\cal R}_3^{\nu\beta},\nonumber\\
	\Re_1\Re_2\Re_3({19})&=&R_1^{\alpha\beta}
	\nabla_\alpha\hat{\cal R}_2^{\mu\nu}\nabla_\beta
	\hat{\cal R}_{3\,\mu\nu},\nonumber\\
	\Re_1\Re_2\Re_3({20})&=&R_1\nabla_\alpha
	\hat{\cal R}_2^{\alpha\mu}\nabla^\beta
	\hat{\cal R}_{3\,\beta\mu}\nonumber,\\
	\Re_1\Re_2\Re_3({21})&=&R_1^{\mu\nu}\nabla_\mu
	\nabla_\lambda\hat{\cal R}_2^{\lambda\alpha}
	\hat{\cal R}_{3\,\alpha\nu},\nonumber\\
	\Re_1\Re_2\Re_3({22})&=&R_1^{\alpha\beta}
	\nabla_\alpha R_2 \nabla_\beta R_3\hat{1},\nonumber\\
	\Re_1\Re_2\Re_3({23})&=&\nabla^\mu R_1^{\nu\alpha}
	\nabla_\nu R_{2\,\mu\alpha}R_3\hat{1},\nonumber\\
	\Re_1\Re_2\Re_3({24})&=&R_1^{\mu\nu}
	\nabla_\mu R_2^{\alpha\beta}
	\nabla_\nu R_{3\,\alpha\beta}
	\hat{1},\nonumber\\
	\Re_1\Re_2\Re_3({25})&=&R_1^{\mu\nu}
	\nabla_\alpha R_{2\,\beta\mu}\nabla^\beta
	R_{3\,\nu}^\alpha\hat{1},
\label{2.9}
	\end{eqnarray}
three structures contain four derivatives
	\begin{eqnarray}
	\Re_1\Re_2\Re_3({26})&=&\nabla_\alpha\nabla_\beta
	R_1^{\mu\nu}\nabla_\mu\nabla_\nu
	R_2^{\alpha\beta}\hat{P}_3,\nonumber\\
	\Re_1\Re_2\Re_3({27})&=&\nabla_\alpha\nabla_\beta
	R_1^{\mu\nu}\nabla_\mu\nabla_\nu R_2^{\alpha\beta}
	R_3\hat{1},\nonumber\\
	\Re_1\Re_2\Re_3({28})&=&\nabla_\mu
	R_1^{\alpha\lambda} \nabla_\nu
	R_{2\,\lambda}^\beta\nabla_\alpha\nabla_\beta
	R_3^{\mu\nu}\hat{1}
\label{2.9a}
	\end{eqnarray}
and one structure contains six derivatives
	\begin{eqnarray}
	\Re_1\Re_2\Re_3({29})&=&\nabla_\lambda\nabla_\sigma
	R_1^{\alpha\beta}\nabla_\alpha\nabla_\beta
	R_2^{\mu\nu}\nabla_\mu\nabla_
	\nu R_3^{\lambda\sigma}\hat{1}.
\label{2.10}
	\end{eqnarray}

Here we present only those elements of the full basis of third-order
invariants
explicitly built in Ref.\onlinecite{JMP1},  whose coefficients are
nonvanishing
in the one-loop effective action.
It turns out that these basis elements are also not completely
independent, and
there exists a nonlocal identity \cite{IV,JMP1}
	\begin{eqnarray}
	&&\int\! d^4 x\, g^{1/2}\, {\rm tr}\,
	{\cal F}^{\rm sym}(\Box_1,\Box_2,\Box_3)
	\left\{-\frac1{48}({\Box_1}^2+{\Box_2}^2
	+{\Box_3}^2)\Re_1\Re_2\Re_3({9})\right.
	\nonumber\\
	&&\ \ \ \ -\frac1{12}({\Box_1}^2+{\Box_2}^2+{\Box_3}^2
 	-2{\Box_1}{\Box_2}-2{\Box_2}{\Box_3}
	-2{\Box_1}{\Box_3})\Re_1\Re_2\Re_3({10})
	\nonumber\\
	&&\ \ \ \ -\frac18{\Box_3}({\Box_1}
	+{\Box_2}-{\Box_3})\Re_1\Re_2\Re_3({11})
	+\frac18(3{\Box_1}+{\Box_2}+
	{\Box_3})\Re_1\Re_2\Re_3({22})
	\nonumber\\
	&&\ \ \ \ -\frac12{\Box_3}\Re_1\Re_2\Re_3({23})
	-\frac12{\Box_1}\Re_1\Re_2\Re_3({24})
	-\frac12({\Box_2}+{\Box_3}-{\Box_1})\Re_1\Re_2\Re_3({25})
	\nonumber\\
	&&\ \ \ \ \left. +\frac12\Re_1\Re_2\Re_3({27})
	+\Re_1\Re_2\Re_3({28})
	\right\}+{\rm O}[\Re^4]=0,
\label{2.11}
	\end{eqnarray}
valid in four-dimensional asymptotically flat spacetime with an
arbitrary
coefficient function ${\cal F}^{\rm sym}(\Box_1,\Box_2,\Box_3)$
{\em completely symmetric} in its operator box arguments. This
identity will be
used to exclude the completely symmetric (under the permutation of
labels
1,2,3) part of the structure $\Re_1\Re_2\Re_3({28})$.

The operator coefficients in (\ref{2.6}) -- the nonlocal formfactors
of the
effective action -- are obtained in  covariant perturbation theory
\cite{II}
from the analogous nonlocal formfactors in the trace of the heat
kernel ${\rm
Tr}\,K(s)={\rm Tr}\,{\rm exp}\,(sH)$ by integrating over the proper
time
according to the equation
	\begin{eqnarray}
	\frac 12\,{\rm Tr\,ln}\,H=
	-\frac 12\,\int_0^\infty\frac{ds}s\,{\rm Tr}\,K(s).
\label{2.12}
	\end{eqnarray}
The trace of the heat kernel calculated to third order in the
curvature in
Ref.\onlinecite{IV}
	\begin{eqnarray}
	{\rm Tr}\,K(s)\!\!\!&=&
	\frac1{(4\pi s)^\omega}\int dx\, g^{1/2}\, \,{\rm tr}\,
	\Big\{\,\hat{1}+s\hat{P}+s^2\sum^{5}_{i=1}f_{i}(-s\Box_2)\,
	\Re_1\Re_2({i})\nonumber\\&&
	+s^3\sum^{11}_{i=1}F_{i}(-s\Box_1,-s\Box_2,-s\Box_3)\,
	\Re_1\Re_2\Re_3({i})\nonumber\\&&
	+s^4\sum^{25}_{i=12}F_{i}(-s\Box_1,-s\Box_2,-s\Box_3)\,
	\Re_1\Re_2\Re_3({i})\nonumber\\&&
	+s^5\sum^{28}_{i=26}F_{i}(-s\Box_1,-s\Box_2,-s\Box_3)\,
	\Re_1\Re_2\Re_3({i})\nonumber\\&&
	+s^6\,\phantom{\sum^{28}_{i=26}}\!\!\!\!\!\!\!\!\!
	F_{29}(-s\Box_1,-s\Box_2,-s\Box_3)\,
	\Re_1\Re_2\Re_3({29})\,
	+{\rm O}[\,\Re^4\,]\,\Big\},             \label{2.13}
	\end{eqnarray}
thus, generates the second-order and the third-order formfactors in
Eq.(\ref{2.6})
	\begin{eqnarray}
	\gamma_i(-\Box_2)=(4\pi)^2\int_0^\infty\frac{ds}s\,
	\frac{s^2}{(4\pi s)^\omega}f_{i}(-s\Box_2),
\label{2.14}
	\end{eqnarray}
	\begin{eqnarray}
	\Gamma_i(-\Box_1,-\Box_2,-\Box_3)=
	(4\pi)^2\int^\infty_0\frac{d s}{s}
	\frac{s^{p_i}}{(4\pi s)^\omega}
	F_i(-s\Box_1,-s\Box_2,-s\Box_3)         \label{2.15}
	\end{eqnarray}
with $p_i=3$ for $i=1$ to 11, $p_i=4$ for $i=12$ to 25, $p_i=5$ for
$i=26$ to
28
and $p_i=6$ for $i=29$. Here $\omega$ plays the role of the parameter
in the
dimensional regularization of ultraviolet divergences at
$\omega\rightarrow 2$.
Simple expressions for  $f_{i}(-s\Box_2)$ obtained in
Ref.\onlinecite{II} give
rise
to second-order form factors
	\begin{eqnarray}
	\gamma_1(-\Box)&=&\frac1{60}
	 \left(-\ln\Big(-\frac{\Box}{\mu^2}\Big)
	+\frac{16}{15}\right),\nonumber\\
	\gamma_2(-\Box)&=&\frac1{180}
 	\left(\ln\Big(-\frac{\Box}{\mu^2}\Big)
	-\frac{37}{30}\right),\nonumber\\
	\gamma_3(-\Box)&=&-\frac1{18},\nonumber\\
	\gamma_4(-\Box)&=&-\frac12\ln
	\Big(-\frac{\Box}{\mu^2}\Big),\nonumber\\
	\gamma_5(-\Box)&=&\frac1{12}
	\left(-\ln\Big(-\frac{\Box}{\mu^2}\Big)+\frac23\right),
\label{2.16}
	\end{eqnarray}
where the parameter $\mu^2>0$ accounts for the ultraviolet
arbitrariness in all the form factors except $\gamma_3(-\Box)$ which
is local
and independent of $\mu^2$ (see Ref.\onlinecite{II}). The
arbitrariness in
$\mu^2$ results from
subtracting (by renormalization) the logarithmic divergences of the
effective
action, accumulating the pole parts of the divergent integrals
(\ref{2.14}).
They are given by the integrated DeWitt coefficient $a_2(x,x)$ at
coincident
points \cite{DW,PhysRep}
	\begin{eqnarray}
	-W^{\rm div}&=&\left(\frac1{2-\omega}+{\rm ln}\,4\pi-{\rm
C}+2\right)\,
	\int d^4x\,g^{1/2}{\rm tr}\,\hat a_2(x,x),
	\;\;\;\omega\rightarrow 2,
\label{2.17}\\
	\hat{a}_2(x,x) &=&
	\frac1{180}\,(R_{\alpha\beta\gamma\delta}
	R^{\alpha\beta\gamma\delta}
	-R_{\mu\nu} R^{\mu\nu})\,\hat{1} \nonumber\\
	&&\qquad\qquad\qquad
	+\frac1{12}\hat{\cal R}_{\mu\nu}\hat{\cal R}^{\mu\nu}
	+\frac12\hat{P}^2+\frac16\Box\hat{P}
	+\frac1{180}\Box R\hat{1},
\label{2.18}
	\end{eqnarray}
which absorbs all the divergences of (\ref{2.14}) on account of the
Gauss-Bonnet identity
	\begin{eqnarray}
	\int\! d^4x\, g^{1/2}\, \,\Big(R_{\alpha\beta\gamma\delta}
	R^{\alpha\beta\gamma\delta}
	-4 R_{\mu\nu} R^{\mu\nu} + R^2\Big)=0.
\label{2.19}
	\end{eqnarray}

In contrast to (\ref{2.16}), the third-order form factors contain no
arbitrary
parameters and are finite. This goes as follows. The integrals
(\ref{2.15}) are
generally divergent, but the divergences cancel in the sum
	\begin{eqnarray}
	\sum_i \Gamma_i(-\Box_1,-\Box_2,-\Box_3)\,
	\Re_1\Re_2\Re_3({i}).
\label{2.20}
	\end{eqnarray}
The mechanism of cancellation is the identity (\ref{2.11}). If, in
this
identity, one puts
	\begin{eqnarray}
	{\cal F}^{\rm sym}(\Box_1,\Box_2,\Box_3) &=&
	-\frac2{45}\frac1{\Box_1\Box_2\Box_3}
	\left(\frac1{2-\omega}+\ln4\pi-{\rm C}
	-\frac13\sum^3_{n=1}\ln(-\Box_n)\right),
\label{2.21}
	\end{eqnarray}
then its left-hand side will be precisely the divergent term of
(\ref{2.20}).
The reason why these divergences appear at all is the
presence of the Riemann tensor in the DeWitt
coefficient (\ref{2.18}) which governs the ultraviolet
divergences in four dimensions.
Since covariant perturbation theory expands the Riemann
tensor in an infinite series in powers of the Ricci
tensor (see Refs. \onlinecite{II,IV,JMP1} and Eq.(\ref{2.3a}) above),
the
divergent
term with the Riemann tensor brings divergent
contributions to the third and all higher orders in the
Ricci curvature. The problem vanishes, however, if one
takes into account the Gauss-Bonnet identity (\ref{2.19}) which,
in four dimensions, eliminates the Riemann tensor
from the (integrated) $a_2(x,x)$. Automatically
eliminated then are also all divergent contributions
of higher orders in the curvature. At each order,
there exists a nonlocal constraint which ensures
this elimination. The hierarchy of these constraints
is generated by the expansion of the Gauss-Bonnet
invariant (see Eq. (6.36) of Ref. \onlinecite{JMP1} and detailed
derivation
in the report \onlinecite{IV}).

It is worth emphasizing the difference between covariant perturbation
theory
\cite{II,IV} and the usual flat-space perturbation theory. In the
latter,
non-covariant vertices {\em of all orders} are divergent. The
divergences,
therefore, proliferate in an apparently uncontrollable way and the
fact that
they cancel in some effect (see, e.g. Ref.\onlinecite{BerGast}) is
usually
regarded as a miracle. In essence, no miracle happens. The physical
effects are
covariant and are determined by covariant vertices, i.e. form factors
of $W$
starting with $N=3$, which are finite. A similar situation takes
place with the
infrared renormalization in quantum electrodynamics \cite{OstrV}.
Covariant
renormalization theory for gauge fields was pioneered by DeWitt
\cite{DW}. For
its extension beyond the one-loop level see Ref.\onlinecite{BV:2l}.

The third-order form factors
	$\Gamma_{i}(-\Box_1,-\Box_2,-\Box_3),\, i=1\ {\rm to}\,29$,
when calculated by Eq.(\ref{2.15}) from the form factors of the heat
kernel,
all linearly express on account of the identity (\ref{2.11}) through
the basic
third-order form factor
	\begin{eqnarray}
	\Gamma(-\Box_1,-\Box_2,-\Box_3)=\int_{\alpha\geq 0}
	d^3\,\alpha\,\frac{\delta(1-\alpha_1-\alpha_2-\alpha_3)}
	{-\alpha_1\alpha_2\Box_3
	-\alpha_1\alpha_3\Box_2
	-\alpha_2\alpha_3\Box_1}     \label{2.22}
	\end{eqnarray}
and the second-order form factors
	\begin{equation}
	{\rm a)\ }\ln({\Box_{n}}/{\Box_{m}}),\ \ \ \
	{\rm b)\ }\frac{\ln(\Box_{n}/\Box_{m})}
	{\Box_{n}-\Box_{m}},\ \ \ \ n,m=1,2,3.     \label{2.23}
	\end{equation}
The coefficients of these expressions are rational
functions of the following general form:
	\begin{equation}
	\frac{P(\Box)}{D^6\Box_1\Box_2\Box_3},        \label{2.24}
	\end{equation}
where $P(\Box)$ is a polynomial, and
	\begin{equation}
	D={\Box_1}^2+{\Box_2}^2+{\Box_3}^2
	-2\Box_1\Box_2-2\Box_1\Box_3-2\Box_2\Box_3.  \label{2.25}
	\end{equation}
The basic third-order form factor $\Gamma(-\Box_1,-\Box_2,-\Box_3)$
is a
solution of the differential equation
	\begin{eqnarray}
	\frac{\partial}{\partial\Box_1}\Gamma &=&
	\frac{\Box_2+\Box_3-\Box_1}D\Gamma
	+\frac1D\ln({\Box_2}/{\Box_1})   \nonumber\\
	&&\,\,\,\,\,\,\,\,\,\,\,\,+\frac1D\ln({\Box_3}/{\Box_1})
	+\frac{\Box_2-\Box_3}{D\Box_1}\ln({\Box_2}/{\Box_3})
\label{2.26}
	\end{eqnarray}
and two other equations, with $\partial/\partial\Box_2$ and
$\partial/\partial\Box_3$, obtained from (\ref{2.26}) by cyclic
symmetry. It
also has the original $\alpha$-representation (\ref{2.22}), Laplace
representation \cite{IV}
	\begin{equation}
	\Gamma(-\Box_1,-\Box_2,-\Box_3)=
	\int^\infty_0\!\! d^3 u\,\frac{\exp(u_1{\Box_1}+
	u_2{\Box_2}+u_3{\Box_3})}
	{u_1u_2+u_2u_3+u_1u_3},
\label{2.27}
	\end{equation}
spectral representation \cite{III,IV}
	\begin{equation}
	\Gamma(-\Box_1,-\Box_2,-\Box_3)= \int_0^\infty {
	{d{m_1}^2\, d{m_2}^2\, d{m_3}^2\, \rho (m_1,m_2,m_3)
	}\over{
	({m_1}^2-\Box_1)({m_2}^2-\Box_2)({m_3}^2-\Box_3)}},
\label{2.28}
	\end{equation}
with the discontinuous spectral weight $\rho (m_1,m_2,m_3)$ as a
function
of masses $m_1= \sqrt{{m_1}^2},\, m_2= \sqrt{{m_2}^2},\,m_3=
\sqrt{{m_3}^2}$
\mathindent = 0pt
\arraycolsep=0pt
	\begin{eqnarray}
	\rho (m_1,m_2,m_3)=\frac{\theta(m_1+m_2-m_3)\,
	\theta(m_1+m_3-m_2)\,\theta(m_3+m_2-m_1)}{\pi
\Big[(m_1\!+m_2\!+m_3\!)(m_1\!+m_2\!-m_3\!)(m_1\!+m_3\!-m_2\!)
	(m_3\!+m_2\!-m_1\!)\Big]^{1/2}},
\label{2.29}
	\end{eqnarray}
\arraycolsep=3pt
\mathindent=\leftmargini
and the generalized spectral representation \cite{IV}, based on the
equation
for Bessel functions
	$\int_0^\infty d y^2J_0(y m_1)J_0(y m_2)
	J_0(y m_3)= 4\rho (m_1,m_2,m_3)$,
	\begin{eqnarray}
	&&\Gamma(-\Box_1,-\Box_2,-\Box_3)=2\int_0^\infty d y^2
	{\cal K}_{0} (y\, \sqrt{-\Box_1})
	{\cal K}_{0} (y\, \sqrt{-\Box_2})
	{\cal K}_{0} (y\, \sqrt{-\Box_3}),          \label{2.30}\\
	&&{\cal K}_{0} (y\, \sqrt{-\Box}) = \frac12
	\int_0^\infty
	\frac{d m^2 J_0(ym)}{{m}^2-\Box}. \label{2.31}
	\end{eqnarray}
The latter is especially important for imposing the boundary
conditions in the
nonlocal form factors of the expectation-value equations in
Lorentzian
spacetime \cite{FrolV,II}.

Explicit expressions for all 29 form factors
$\Gamma_i(-\Box_1,-\Box_2,-\Box_3)$ in terms of
$\Gamma(-\Box_1,-\Box_2,-\Box_3)$ are presented in the report
\onlinecite{IV}.
These expressions take pages and are manageable only in the format of
the
computer algebra program {\em Mathematica} \cite{files}. Therefore,
they will
not be presented here. The $\alpha$-representation, Laplace
representation and
generalized spectral representation are also obtained in this report
for all
form factors $\Gamma_i(-\Box_1,-\Box_2,-\Box_3)$, as well as their
asymptotics
of large and small values of their arguments. The small-$\Box$
behaviours
\cite{CQGas} determine the vacuum radiation effect at the
asymptotically flat
infinity \cite{Armen}.

\section{Derivation of the trace anomaly}
\indent
If the operator $H$ in (\ref{2.1}) corresponds to a conformal
invariant quantum
field in four dimensions, the trace anomaly should follow from
(\ref{2.6})
simply by varying $W$ and taking the trace. Obtaining the correct
trace anomaly
is also a powerful check on the result for $W$. To have as many
currvature
structures as possible involved in the check,  we choose the
following quantum
field model:
	\begin{equation}
	S [ \varphi ] = \frac12\int\! dx\, g^{1/2}\, \,
	\left(\nabla_{\mu}\varphi^{T}
	\nabla^{\mu}\varphi+\frac{R}{6}\varphi^{T}\varphi
	+\frac{\lambda^2}{4!}
	(\varphi^{T}\varphi)^2 \right),              \label{3.1}
	\end{equation}
	\begin{equation}
	\nabla_{\mu}\varphi
	= \partial_{\mu}\varphi + {A}_{\mu}\hat{G}\varphi,\ \ \ \
	\nabla_{\mu}\varphi^{T} =
	\partial_{\mu}\varphi^{T}+A_\mu
	\varphi^{T}\hat{G}^T,                           \label{3.2}
	\end{equation}
	\begin{equation}
	 \varphi=\left(\begin{array}{c}\varphi_1
	\\ \varphi_2
	\end{array} \right),\,\,\,\,\,
	\hat{G}=\left(\begin{array}{cc} 0&1
	\\ -1& 0
	\end{array} \right),
\label{3.3}
	\end{equation}
where (\ref{3.1}) is the euclidean action of the complex scalar
quantum field $\varphi=\varphi_1+i\varphi_2$, rewritten in terms of
the
real components, and $T$ denotes the matrix transposition. The
electromagnetic
and gravitational fields in (\ref{3.1})
are classical.

The action (\ref{3.1}) is invariant under the local
conformal transformations
	\begin{equation}
	\delta_\sigma g^{\mu\nu}(x)=\sigma(x)g^{\mu\nu}(x),\ \ \ \
	\delta_\sigma\varphi(x)=\frac12\sigma(x)\varphi(x),\ \ \ \
	\delta_\sigma A_\mu(x)=0
\label{3.4}
	\end{equation}
with the parameter $\sigma(x)$. The hessian of the action has
the form (\ref{2.1}) (times a local matrix) in which the potential
and the
commutator
curvature are
	\begin{equation}
	\hat{P}=-\frac{2\lambda^2}{4!}
	\left(\begin{array}{cc} 3{\varphi_1}^2
	+{\varphi_2}^2&2{\varphi_1}{\varphi_2}     \label{3.5}
	\\
	2{\varphi_1}{\varphi_2} & 3{\varphi_2}^2
	+{\varphi_1}^2\end{array}\right),
	\end{equation}
	\begin{equation}
	\hat{\cal R}_{\mu\nu}=\hat{G}(
	\partial_\mu A_\nu-
	\partial_\nu A_\mu)                        \label{3.6}
	\end{equation}
with $\hat{G}$ in (\ref{3.3}).

{}From (\ref{3.4})--(\ref{3.6}) we find the conformal
transformation laws for the curvatures and
$\Box$-operators:
	\begin{eqnarray}
	&&\delta_\sigma \hat{P}=\sigma\hat{P},\ \ \ \
	\delta_\sigma\hat{\cal R}_{\mu\nu}=0, \nonumber
	\\
	&&\delta_\sigma R_{\mu\nu}=
	(\nabla_\mu\nabla_\nu+\frac12g_{\mu\nu}\Box)\sigma,
	\,\,\,\,\delta_\sigma R=(3\Box+R)\sigma,      \nonumber
	\\
	&&(\delta_\sigma\Box)\hat{P}=\sigma\Box\hat{P}
	 -\nabla_\alpha\sigma\nabla^\alpha\hat{P},   \nonumber
	\\
	&&(\delta_\sigma\Box)\hat{\cal R}_{\mu\nu} =
	\sigma\Box\hat{\cal R}_{\mu\nu}
	+\hat{\cal R}_{\mu\nu}\Box\sigma
	+\nabla_\mu\sigma\nabla^\alpha\hat{\cal R}_{\alpha\nu}
	-\nabla_\nu\sigma\nabla^\alpha\hat{\cal R}_{\alpha\mu}\, ,
\nonumber
	\\
	&&(\delta_\sigma\Box)R_{\mu\nu} =
	\sigma\Box R_{\mu\nu}
	+R_{\mu\nu}\Box\sigma
	+\nabla_\alpha\sigma\nabla^\alpha R_{\mu\nu}
	+\nabla_{(\mu}\sigma\nabla_{\nu)} R
	-2\nabla^\alpha\sigma\nabla_{(\mu}R_{\nu)\alpha}\, ,
\nonumber
	\\
	&&(\delta_\sigma\Box)R=\sigma\Box R-
	\nabla_\alpha\sigma \nabla^\alpha R.        \label{3.6a}
	\end{eqnarray}
Having got these laws, one may already forget the
particular content of the model, and merely consider
the transformation (\ref{3.6a}) in the
effective action. For the dimensionally regularized
\cite{f1} one-loop effective action (6.1), the result should be
exactly
	\begin{equation}
	-\delta_\sigma W=-\frac1{2(4\pi)^2}\int\! dx\, g^{1/2}\,
	\sigma(x)\,{\rm tr}\,\hat{a}_2(x,x),
\label{3.7}
	\end{equation}
where $\hat{a}_2(x,x)$ is the second DeWitt coefficient at coincident
points
(\ref{2.18})
\cite{f2}. Expression (\ref{3.7}) is the general form of the
conformal anomaly
in four
dimensions \cite{DDI,Duff,Brown,FV-conf}. For the model above,
	\begin{equation}
	\delta_\sigma=\int dx\,\left(
	\sigma(x)g^{\mu\nu}
	\frac{\delta}{\delta g^{\mu\nu}}
	+\frac12\sigma(x)\varphi
	\frac{\delta}{\delta\varphi}\right),          \label{3.9}
	\end{equation}
and
	\begin{equation}
	g^{-1/2}\left(g^{\mu\nu}
	\frac{\delta W}{\delta g^{\mu\nu}}
	+\frac12\varphi
	\frac{\delta W}{\delta\varphi}\right) =
	\frac1{2(4\pi)^2}{\rm tr}\,\hat{a}_2(x,x).
\label{3.10}
	\end{equation}

In the present technique, Eq. (\ref{3.7}) with $\hat{a}_2(x,x)$ given
by
(\ref{2.18}) can be obtained only with a given accuracy ${\rm
O}[\Re^n]$
and with the Riemann tensor expressed through the
Ricci tensor. To lowest order, one may use the
expression for $R^2_{\alpha\beta\mu\nu}$
given in Appendix A of paper \onlinecite{II}. After elimination
of the Riemann tensor from (\ref{2.18}), Eq. (\ref{3.7})
takes the form
	\begin{eqnarray}
	-\delta_\sigma W &=&
	{\frac1{2(4\pi)^2}\int\! dx\, g^{1/2}\, \,{\rm tr}\,}
	\left\{-\frac16(\Box\hat{P})\sigma
	\right.\nonumber
	\\&&
	-\frac1{180}(\Box R)\sigma\hat{1}
	-\frac1{12}\hat{\cal R}^2_{\mu\nu}\sigma
	-\frac12\hat{P}^2\sigma\nonumber\\&&
	-\frac1{180}
	\left(1+2\frac{\Box_1}{\Box_2}
	-4\frac{\Box_3}{\Box_1}+\frac{{\Box_3}^2}
	{\Box_1\Box_2}\right)
	{R_1^{\mu\nu}R_{2\mu\nu}\sigma_3\hat{1}}
	\nonumber\\
	&&-\frac1{45}\left(\frac2{\Box_1}-\frac{\Box_3}
	{\Box_1\Box_2}\right)
	{\nabla^\mu R^{\nu\lambda}_1
         \nabla_\nu R_{2\mu\lambda}\sigma_3\hat{1}}
	\nonumber\\&&\left.
	-\frac1{45}\frac1{\Box_1\Box_2}
	{\nabla_\alpha\nabla_\beta R_{1\mu\nu}
         \nabla^\mu\nabla^\nu
	R_2^{\alpha\beta}\sigma_3\hat{1}}\right\}+{\rm O}[\Re^3]
\label{3.11}
	\end{eqnarray}
where the notation in the nonlocal terms is the
same as before with $\sigma$ playing the role of
the third curvature. It is the latter equation
that will be checked below by a direct calculation
with $W$ in (\ref{2.6}).

We begin this check with calculating the
result of the transformation (\ref{3.6a})
in the quadratic terms of $W$. For the quadratic
terms of (\ref{2.6}) we have
	\begin{eqnarray}&&
	\delta_\sigma\int\! dx\, g^{1/2}\, {\rm tr}\left\{
	\sum^5_{i=1}\gamma_i(-\Box_2)\Re_1\Re_2({i})\right\}
	=\int\! dx\, g^{1/2}\, {\rm tr}\left\{
	-\frac3{18}\hat{P}\Box\sigma \right.\nonumber
	\\&&\qquad\quad
	+\frac1{30}R^{\mu\nu}\left[\gamma(-\Box)
	+\frac{16}
	{15}\right]\Big(\nabla_\mu\nabla_\nu\sigma
	+\frac12g_{\mu\nu}\Box\sigma\Big)\hat{1} \nonumber
	\\&&\qquad\quad
	-\frac3{90}R\left[\gamma(-\Box)+\frac{37}
	{30}\right]
	\Box\sigma\hat{1} \nonumber
	\\&&\qquad\quad
	+\frac1{12}
	\hat{\cal R}_{\mu\nu}
	\Big(\delta_\sigma\gamma(-\Box)\Big)
	\hat{\cal R}^{\mu\nu}
	+\frac1{2}
	\hat{P}
	\Big(\delta_\sigma\gamma(-\Box)\Big)
	\hat{P} \nonumber\\&&\qquad\quad
	\left.
	+\frac1{60}
	R_{\mu\nu}
	\Big(\delta_\sigma\gamma(-\Box)\Big)
	R^{\mu\nu}\hat{1}
	-\frac1{180}R
	\Big(\delta_\sigma\gamma(-\Box)\Big)
	R \hat{1}\right\}
\label{3.12}
	\end{eqnarray}
where
	\begin{equation}
	\gamma(-\Box)=-\ln\left(-\frac{\Box}
	{\mu^2}\right).
\label{3.13}
	\end{equation}
In the term linear in $R^{\mu\nu}$, for being able
to use the Bianchi identity, one must commute
$\gamma(-\Box)$ with $\nabla_\mu\nabla_\nu$, and the
commutator cannot be neglected. As a result of
this commutation, the linear nonlocal terms cancel,
and we obtain
\begin{eqnarray}&&
\delta_\sigma\int\! dx\, g^{1/2}\, {\rm tr}\left\{
\sum^5_{i=1}\gamma_i(-\Box_2)\Re_1\Re_2({i})\right\}
\nonumber\\&&\ \ \ \ \ \ \ \
=\int\! dx\, g^{1/2}\, {\rm tr}\left\{\right.
-\frac1{6}(\Box\hat{P})\sigma
-\frac1{180}(\Box R)\sigma\hat{1} \nonumber
\\&&\ \ \ \ \ \ \ \ \ \ \ \
+\frac1{30}R_{\mu\nu}\Big[\gamma(-\Box),
\nabla^\mu\nabla^\nu\Big]\sigma\hat{1} \nonumber
\\&&\ \ \ \ \ \ \ \ \ \ \ \
+\frac1{12}
\hat{\cal R}_{\mu\nu}
\Big(\delta_\sigma\gamma(-\Box)\Big)
\hat{\cal R}^{\mu\nu}
+\frac1{2}
\hat{P}
\Big(\delta_\sigma\gamma(-\Box)\Big)
\hat{P} \nonumber\\&&\ \ \ \ \ \ \ \ \ \ \ \
\left.
+\frac1{60}
R_{\mu\nu}
\Big(\delta_\sigma\gamma(-\Box)\Big)
R^{\mu\nu}\hat{1}
-\frac1{180}
R
\Big(\delta_\sigma\gamma(-\Box)\Big)
R \hat{1}\right\}                              \label{3.15}
\end{eqnarray}
where the first two terms correctly reproduce the
linear contributions to the anomaly \cite{II}, and the
remaining terms are already quadratic in the
curvature.

For the calculation of the quadratic terms in
(\ref{3.15}) we use the spectral representation
	\begin{equation}
	\gamma(-\Box) =
	\int^\infty_0dm^2\,
	\Big(\frac1{m^2-\Box}-\frac1{m^2+\mu^2}\Big)
	\end{equation}
and the commutation (variation) rule for the inverse operator
	\begin{equation}
	\Big[\frac1{m^2-\Box},\nabla^\mu\nabla^\nu\Big] = -
	\frac1{m^2-\Box}
	\Big[-\Box,\nabla^\mu\nabla^\nu\Big]
	\frac1{m^2-\Box},
\label{3.16}
	\end{equation}
	\begin{equation}
	\delta_\sigma\frac1{m^2-\Box} = -
	\frac1{m^2-\Box}\delta_\sigma(-\Box)
	\frac1{m^2-\Box}
\label{3.17}
	\end{equation}
where, within the required accuracy, the
factors on the right-hand sides can already be
commuted freely. Doing the spectral-mass integral
then gives
	\begin{eqnarray}&&
	\int\! dx\, g^{1/2}\,
	R_{\mu\nu}[\gamma(-\Box),
	\nabla^\mu\nabla^\nu]\sigma \nonumber\\&&\ \ \ \
	=-\int\! dx\, g^{1/2}\, \frac{\ln(\Box_1/\Box_3)}
	{(\Box_1-\Box_3)}
	[\Box_3,\nabla^\mu_3\nabla^\nu_3]
	R_{1\mu\nu}\sigma_3
	+{\rm O}[R^3..],
\label{3.18}
	\end{eqnarray}
and, similarly,
	\begin{eqnarray}&&
	\int\! dx\, g^{1/2}\, {\rm tr}\,
	\Re_1\Big(\delta_\sigma\gamma(-\Box_2)\Big)\Re_2 \nonumber
	\\&&\ \ \ \ \ \ \ \
	=-\int\! dx\, g^{1/2}\, {\rm tr}\,\frac{\ln(\Box_1/\Box_2)}
	{(\Box_1-\Box_2)}
	(\delta_\sigma\Box_2)\Re_1\Re_2
	+{\rm O}[\Re^3].
\label{3.20}
	\end{eqnarray}
There remain to be used in (\ref{3.20}) the
transformation laws (\ref{3.6a}), and
in (\ref{3.18}) the expression for the commutator
	\begin{eqnarray}
	[\Box,\nabla_\mu\nabla_\nu]\sigma &=&
	2\nabla_{(\mu}R_{\nu)\alpha}\nabla^\alpha\sigma
	+2R_{\alpha(\mu}\nabla_{\nu)}\nabla^\alpha\sigma \nonumber
	\\&&\,\,\,\,\,\,\,\,\,\,\,\,\,
	-\nabla_\alpha R_{\mu\nu}\nabla^\alpha\sigma
	-2R_{\alpha\nu\beta\mu}
	\nabla^\alpha\nabla^\beta\sigma           \label{3.21}
	\end{eqnarray}
in which the Riemann tensor should be expressed through the Ricci
tensor
by Eq.(\ref{2.3a}).

The final result for (\ref{3.15}) is
	\begin{eqnarray}&&
	\delta_\sigma\int\! dx\, g^{1/2}\, {\rm tr}\left\{
	\sum^5_{i=1}\gamma_i(-\Box_2)\Re_1\Re_2({i})\right\}
\nonumber
	\\&&\ \ \ \ \ \ \ \
	=\int\! dx\, g^{1/2}\, {\rm tr}\left\{
	-\frac1{6}(\Box\hat{P})\sigma
	-\frac1{180}(\Box R)\sigma\hat{1}
	\right.
	\nonumber\\&&\ \ \ \ \ \ \ \ \ \ \ \ \ \ \ \
	\left.
	+\sum^{10}_{i=1}M_i(\Box_1,\Box_2,\Box_3)
	\Re_1\Re_2\sigma_3({i})\right\}+{\rm O}[\Re^3],
\label{3.22}
	\end{eqnarray}
where $\Re_1\Re_2\sigma_3({i})$ are the following ten tensor
structures
\begin{eqnarray}
	\Re_1\Re_2\sigma_3({1}) &=&
	{R_1R_2\sigma_3\hat{1}},  \nonumber
	\\
	\Re_1\Re_2\sigma_3({2}) &=&
	{R_1^{\mu\nu}R_{2\mu\nu}\sigma_3\hat{1}}, \nonumber
	 \\
	\Re_1\Re_2\sigma_3({3}) &=&
	{R_1^{\mu\nu}\nabla_\nu\nabla_\mu R_2\sigma_3\hat{1}},
\nonumber
	\\
	\Re_1\Re_2\sigma_3({4}) &=&
	{\nabla^\mu R^{\nu\lambda}_1
         \nabla_\nu R_{2\mu\lambda}\sigma_3\hat{1}},  \nonumber
	\\
	\Re_1\Re_2\sigma_3({5}) &=&
	{\nabla_\alpha\nabla_\beta R_{1\mu\nu}
	\nabla^\mu\nabla^\nu R_2^{\alpha\beta}\sigma_3\hat{1}},
\nonumber
	\\
	\Re_1\Re_2\sigma_3({6}) &=&
	{\hat{P}_1 R_2\sigma_3},      	\nonumber
	\\
	\Re_1\Re_2\sigma_3({7}) &=&
	{\nabla_\alpha\nabla_\beta\hat{P}_1
	R_2^{\alpha\beta}\sigma_3},      \nonumber
	\\
	\Re_1\Re_2\sigma_3({8}) &=&
	{\hat{P}_1\hat{P}_2\sigma_3},   \nonumber
	\\
	\Re_1\Re_2\sigma_3({9}) &=&
	{\hat{\cal R}_1^{\mu\nu}\hat{\cal R}_{2\mu\nu}\sigma_3},
\nonumber
	\\
	\Re_1\Re_2\sigma_3({10}) &=&
	{\nabla_\mu\hat{\cal R}_1^{\mu\alpha}\nabla^\nu
	\hat{\cal R}_{2\nu\alpha}\sigma_3},
\label{3.23}
	\end{eqnarray}
and for the form factors $M_i(\Box_1,\Box_2,\Box_3)$ we obtain
	\begin{eqnarray}
	&&M_1 =\frac{(-2\Box_1+5\Box_3)}{720}\,
	\frac{\ln(\Box_1/\Box_2)}
	{(\Box_1-\Box_2)}
	+\frac{(\Box_1+\Box_2-\Box_3)}{120}\,
	\frac{\ln(\Box_1/\Box_3)}
	{(\Box_1-\Box_3)},        \nonumber
	\\
	&&M_2 =\frac{(-2\Box_1-\Box_3)}{120}\,
	\frac{\ln(\Box_1/\Box_2)}{(\Box_1-\Box_2)}
	+\frac{(\Box_1-\Box_2-\Box_3)
	(\Box_1-\Box_3)}{60\Box_2}\,
	\frac{\ln(\Box_1/\Box_3)}{(\Box_1-\Box_3)},    \nonumber
	\\
	&&M_3 = -\frac1{30}\,
	\frac{\ln(\Box_1/\Box_3)}{(\Box_1-\Box_3)}-\frac1{30}\,
	\frac{\ln(\Box_2/\Box_3)}{(\Box_2-\Box_3)},   \nonumber	\\
	&&M_4 = -\frac1{30}\,
	\frac{\ln(\Box_1/\Box_2)}{(\Box_1-\Box_2)}
	+\frac{(\Box_1-\Box_2-\Box_3)}{15\Box_2}\,
	\frac{\ln(\Box_1/\Box_3)}{(\Box_1-\Box_3)}, \nonumber
	\\
	&&M_5 = \frac1{15\Box_2}\,
	\frac{\ln(\Box_1/\Box_3)}{(\Box_1-\Box_3)},\,\,\,\,
	M_6 = 0,\,\,\,\,M_7 = 0,       \nonumber
	\\
	&&M_8 = \frac{(-2\Box_2-\Box_3)}4\,\frac{\ln(\Box_1/\Box_2)}
	{(\Box_1-\Box_2)},\,\,\,\,
	M_9 = -\frac{\Box_3}{12}\,
	\frac{\ln(\Box_1/\Box_2)}{(\Box_1-\Box_2)}, \nonumber
	\\
	&&M_{10} = \frac16\,\frac{\ln(\Box_1/\Box_2)}
	{(\Box_1-\Box_2)}.
\label{3.24}
	\end{eqnarray}

The conformal transformation in the cubic terms
of the effective action (\ref{2.6}) is easier to carry
out because, within the required accuracy, only the
curvatures in $\Re_1\Re_2\Re_3$ need be varied.
The result is again a sum of contributions of the
ten tensor structures (\ref{3.23}):
\mathindent=0pt
	\begin{eqnarray}&&
	\delta_\sigma\int\! dx\, g^{1/2}\, {\rm tr}\left\{
	\sum^{29}_{i=1}\Gamma_{i}(-\Box_1,-\Box_2,-\Box_3)
	\Re_1\Re_2\Re_3({i})\right\}  \nonumber
	\\
	&&\ \ \ \,\,\,\,\,
	=\int\! dx\, g^{1/2}\, {\rm tr}\left\{
	\sum^{10}_{i=1}
	N_i(\Box_1,\Box_2,\Box_3)\Re_1
	\Re_2\sigma_3({i})\right\}
	+{\rm O}[\Re^3],                                \label{3.25}
	\end{eqnarray}
where the form factors $N_i(\Box_1,\Box_2,\Box_3)$
are the following combinations of the form factors
$\Gamma_{i}(-\Box_1,-\Box_2,-\Box_3)$:
	\begin{eqnarray} N_{1} &=& (-\Box_1 + 2\Box_3)
	{\Gamma_{11}}
	{\big|}_{\Box_1\leftrightarrow\Box_3}+
	\frac14\Big((\Box_1 - \Box_2 - \Box_3)
	(-\Box_1 + \Box_2 - \Box_3)    \nonumber
	\\
	&&\ \ \ \ + \Box_3(-\Box_1 - \Box_2 + \Box_3)\Big)
	{\Gamma_{22}}
	{\big|}_{\Box_1\leftrightarrow\Box_3}
	\nonumber
	\\
	&& +\left(-\frac14(\Box_1 - \Box_2 - \Box_3)^2 -
\frac12\Box_2\Box_3
	\right.+ \frac14(\Box_1 - \Box_2 - \Box_3)
	(-\Box_1 - \Box_2 + \Box_3)\nonumber
	\\&&\ \ \ \ \left. -
	\frac14\Box_3(-\Box_1 - \Box_2 + \Box_3)\right)
	{\Gamma_{23}}
	{\big|}_{\Box_1\leftrightarrow\Box_3}\nonumber
	\\&& +
	\frac1{16}(-\Box_1 - \Box_2 + \Box_3)^2{\Gamma_{25}}
	{\big|}_{\Box_1\leftrightarrow\Box_3} \nonumber
	\\
	&& +\frac12(-\Box_1 + \Box_3)\left(\frac12(\Box_1 - \Box_2 -
\Box_3)^2
	+ \Box_2\Box_3\right){\Gamma_{27}}
	{\big|}_{\Box_1\leftrightarrow\Box_3}\nonumber
	\\
	&& +\frac1{16}(-\Box_1 + \Box_2 - \Box_3)
	(-\Box_1 - \Box_2 + \Box_3)^2
	{\Gamma_{28}}{\big|}_{\Box_1\leftrightarrow\Box_3}\nonumber
	\\&&
	+ 9\Box_3\Gamma_{9} +
	\frac38(-\Box_1 - \Box_2 + \Box_3)\Gamma_{10}
	-\frac32\Box_3(-\Box_1 - \Box_2 + \Box_3)\Gamma_{22}\nonumber
	\\
	&&+ \frac14(\Box_1 - 2\Box_3)(-\Box_1 - \Box_2 + \Box_3)
	\Gamma_{24}\nonumber
	\\
	&& +\frac18(\Box_1 - \Box_2 - \Box_3)(-\Box_1 - \Box_2 +
\Box_3)
	\Gamma_{25} \nonumber
	\\&& +
	\frac1{32}(-\Box_1 - \Box_2 + \Box_3)
	\Big((\Box_1 - \Box_2 - \Box_3)(-\Box_1 + \Box_2 - \Box_3)
	\nonumber
	\\
	&&\ \ \ \  + \Box_3(-\Box_1 - \Box_2 + \Box_3)\Big)
	\Gamma_{28},   \nonumber\\
	N_{2} &=& \frac14\Big((\Box_1 - \Box_2 - \Box_3)
	(-\Box_1 + \Box_2 - \Box_3)+ \Box_3(-\Box_1 - \Box_2 +
\Box_3)\Big)
	{\Gamma_{24}}{\big|}_{\Box_1\leftrightarrow\Box_3}\nonumber
	\\
	&& + \frac32\Box_3\Gamma_{10}
	+ 3\Box_3\Gamma_{11},   \nonumber
	\\
	N_{3} &=& -\frac12(-\Box_1 - \Box_2 + \Box_3)
	{\Gamma_{24}}{\big|}_{\Box_1\leftrightarrow\Box_2}
	+\frac12 (-\Box_1 + \Box_2 - \Box_3)
	{\Gamma_{25}}{\big|}_{\Box_1\leftrightarrow\Box_2}\nonumber
	\\&&
     + \frac12(-\Box_1 - \Box_2 + \Box_3)
	{\Gamma_{25}}{\big|}_{\Box_1\leftrightarrow\Box_3}\nonumber
	\\&& +
	\frac14(-\Box_1 + \Box_2 - \Box_3)(-\Box_1 - \Box_2 + \Box_3)
	{\Gamma_{28}}{\big|}_{\Box_1\leftrightarrow\Box_3}\nonumber
	\\&&
     + 2{\Gamma_{11}}{\big|}_{\Box_2\leftrightarrow\Box_3} +
	(-\Box_1 + \Box_2 - \Box_3)
	{\Gamma_{23}}{\big|}_{\Box_2\leftrightarrow\Box_3}
	\nonumber
	\\&& +
	\left(\frac12(-\Box_1 + \Box_2 - \Box_3)^2 +
\Box_1\Box_3\right)
	{\Gamma_{27}}{\big|}_{\Box_2\leftrightarrow\Box_3}\nonumber
	\\&& +
	\frac14\Big((\Box_1 - \Box_2 - \Box_3)
	(-\Box_1 - \Box_2 + \Box_3)\nonumber
         - \Box_3(-\Box_1 - \Box_2 + \Box_3)\Big)
	{\Gamma_{28}}{\big|}_{\Box_2\leftrightarrow\Box_3}\nonumber
	\\&& +
         3\Gamma_{10} - 6\Box_3\Gamma_{22}
	+ (\Box_1 - 2\Box_3)\Gamma_{24}
	+\frac12(\Box_1 - \Box_2 - 2\Box_3)\Gamma_{25}\nonumber
	\\&& +
	\frac14\Big((\Box_1 - \Box_2 - \Box_3)
	(-\Box_1 + \Box_2 - \Box_3)+ \Box_3(-\Box_1 - \Box_2
	+ \Box_3)\Big)\Gamma_{28}\nonumber
	\\&& +
	\frac38(-\Box_1 + \Box_3)\Big((-\Box_1 + \Box_2 - \Box_3)^2
	+ 2\Box_1\Box_3\Big)\Gamma_{29}, \nonumber
	\\
	N_{4} &=& \frac12\Box_3{\Gamma_{25}}
	{\big|}_{\Box_1\leftrightarrow\Box_3}
	+ 3\Gamma_{10} + 3\Box_3\Gamma_{23} +
	(\Box_1 - \Box_2 - 2\Box_3)\Gamma_{25}\nonumber
	\\&&
     + \frac14\Big((\Box_1 - \Box_2 - \Box_3)(-\Box_1 + \Box_2
	- \Box_3)\Box_3(-\Box_1 - \Box_2 + \Box_3)\Big)
	\Gamma_{28}, \nonumber
	\\
	N_{5} &=& {\Gamma_{25}}
	{\big|}_{\Box_1\leftrightarrow\Box_3} - (\Box_1 - \Box_2 +
2\Box_3)
	{\Gamma_{28}}{\big|}_{\Box_1\leftrightarrow\Box_3}
	- 2\Gamma_{24} +3\Box_3\Gamma_{27}\nonumber
	\\&&
	+ \frac34\Big((-\Box_1 + \Box_2 - \Box_3)^2 +
2\Box_1\Box_3\Big)
	\Gamma_{29}, \nonumber
	\\
	N_{6} &=& -\frac34\Box_3(-\Box_1 - \Box_2 + \Box_3)
	{\Gamma_{15}}\big|_{\Box_1\rightarrow\Box_2,
	\Box_2\rightarrow\Box_3,\Box_3\rightarrow\Box_1}\nonumber
	\\&&
	+ 6\Box_3{\Gamma_{4}}{\big|}_{\Box_1\leftrightarrow\Box_3} +
    	(-\Box_1 + 2\Box_3)
	{\Gamma_{5}}{\big|}_{\Box_1\leftrightarrow\Box_3}\nonumber
	\\&&
  	+ \frac14\Big((\Box_1 - \Box_2 - \Box_3)
	(-\Box_1 + \Box_2 - \Box_3)+\Box_3(-\Box_1 - \Box_2 +
\Box_3)\Big)
	{\Gamma_{15}}{\big|}_{\Box_1\leftrightarrow\Box_3} \nonumber
	\\&& +
    	\left(\frac14(\Box_1 - \Box_2 - \Box_3)\Box_3 	+
\frac12(\Box_1 - \Box_2 -
\Box_3) (-\Box_1 + \Box_3)\right)
	{\Gamma_{16}}{\big|}_{\Box_1\leftrightarrow\Box_3}\nonumber
	\\&& +
    	\frac12(-\Box_1 + \Box_3)\!\left(\frac12
	(\Box_1 - \Box_2 - \Box_3)^2
      	+ \Box_2\Box_3\right)\!{\Gamma_{26}}
	{\big|}_{\Box_1\leftrightarrow\Box_3},    \nonumber
	\\
	N_{7} &=&
-3\Box_3{\Gamma_{15}}\big|_{\Box_1\rightarrow\Box_2,
	\Box_2\rightarrow\Box_3,\Box_3\rightarrow\Box_1}
	+ 2{\Gamma_{5}}
	{\big|}_{\Box_1\leftrightarrow\Box_3}+(\Box_1 - \Box_2 -
\Box_3)
	{\Gamma_{16}}{\big|}_{\Box_1\leftrightarrow\Box_3}\nonumber
	\\&&
 	+\left(\frac12(\Box_1 - \Box_2 - \Box_3)^2+
\Box_2\Box_3\right)
	{\Gamma_{26}}{\big|}_{\Box_1\leftrightarrow\Box_3},\nonumber
	\\
	N_{8} &=& \left(\frac14(\Box_1 - \Box_2 - \Box_3)^2
  	+ \frac12\Box_2\Box_3\right)
	{\Gamma_{17}}
	{\big|}_{\Box_1\leftrightarrow\Box_3}
  	+ 3\Box_3\Gamma_{6}, \nonumber
	\\
	N_{9} &=& 3\Box_3{\Gamma_{7}}
	{\big|}_{\Box_1\leftrightarrow\Box_3}
   	+ \frac14(\Box_1 + \Box_2 + 3\Box_3)
	{\Gamma_{8}}{\big|}_{\Box_1\leftrightarrow\Box_3}-
    	\frac12\Box_1\Box_2{\Gamma_{18}}
	{\big|}_{\Box_1\leftrightarrow\Box_3} \nonumber
	\\&&
     	+ \frac14\Big((\Box_1 - \Box_2 - \Box_3)
	(-\Box_1+ \Box_2 - \Box_3)+ \Box_3(-\Box_1 - \Box_2 +
\Box_3)\Big)
	{\Gamma_{19}}
	{\big|}_{\Box_1\leftrightarrow\Box_3}\nonumber
	\\&& +
    	\frac14\Big(\Box_2(\Box_1 - \Box_2 - \Box_3) -
\Box_2\Box_3\Big)
	{\Gamma_{21}}{\big|}_{\Box_1\leftrightarrow\Box_3}, \nonumber
	\\
	N_{10} &=&{\Gamma_{8}}{\big|}_{\Box_1\leftrightarrow\Box_3} +
	\frac12(-\Box_1 - \Box_2 + 2\Box_3)
	{\Gamma_{18}}{\big|}_{\Box_1\leftrightarrow\Box_3}\nonumber
	\\&& +
	3\Box_3{\Gamma^{\rm sym}_{20}}
	{\big|}_{\Box_1\leftrightarrow\Box_3} +
	\frac12(\Box_1 - \Box_2 - \Box_3)
	{\Gamma_{21}}{\big|}_{\Box_1\leftrightarrow\Box_3}
\label{3.26}
	\end{eqnarray}
\mathindent=\parindent
with ${\Gamma_{11}}{\big|}_{\Box_1\leftrightarrow\Box_3}$, etc.
denoting the
obvious permutations of the box operator arguments of
$\Gamma_{i}(-\Box_1,-\Box_2,-\Box_3)$.

The total result of (\ref{3.22}) and (\ref{3.25}) is the
following conformal variation of the effective action (\ref{2.6}):
	\begin{eqnarray}
	-\delta_\sigma W &=&
	{\frac1{2(4\pi)^2}\int\! dx\, g^{1/2}\, \,{\rm tr}\,}
	\left\{-\frac16(\Box\hat{P})\sigma
	-\frac1{180}(\Box R)\sigma\hat{1} \right.\nonumber
	\\&&\left.
	+\sum^{10}_{i=1}(M_i+N_i)\Re_1\Re_2\sigma_3({i})\right\}
	+{\rm O}[\Re^3],                     \label{3.27}
	\end{eqnarray}
where the quadratic terms are determined by the sum
$M_i+N_i$. There remain to be calculated the linear
combinations of the third-order form factors
$\Gamma_{i}$ in (\ref{3.26}). The most straightforward way of
calculating these
linear combinations is using the explicit expressions for $\Gamma_i$
given
in Ref.\onlinecite{IV}. It is gratifying
to observe that all terms with the basic third-order
form factor $\Gamma(-\Box_1,-\Box_2,-\Box_3)$ cancel in the
combinations $N_i$, all terms with the second-order
form factors $\ln(\Box_n/\Box_m)$ cancel in the
combinations $N_i+M_i$, and there remain only trees:
	\begin{eqnarray}
	&&\frac12\Big(M_{1}+N_{1}+{M_{1}}
	{\big|}_{\Box_1\leftrightarrow\Box_2}+{N_{1}}
	{\big|}_{\Box_1\leftrightarrow\Box_2}\Big) = 0 ,\nonumber
	\\
	&&\frac12\Big(M_{2}+N_{2}+{M_{2}}
	{\big|}_{\Box_1\leftrightarrow\Box_2}+{N_{2}}
	{\big|}_{\Box_1\leftrightarrow\Box_2}\Big)=
 	  -\frac1{180}-\frac{\Box_1}{180\Box_2}-\frac{\Box_2}
	{180\Box_1}\nonumber
	\\
	&&\,\,\,\,\,\,\,\,\,\,\,\,\,\,\,\,\,\,\,\,
	\,\,\,\,\,\,\,\,\,\,\,\,\,\,\,\,\,\,\,\,\,\,\,\,\,\,\,\,
   	+\frac{\Box_3}
	{90\Box_1}+\frac{\Box_3}
	{90\Box_2}
   	-\frac{{\Box_3}^2}
	{180\Box_1\Box_2} , \nonumber
	\\
	&&M_{3}+N_{3} = 0 ,\nonumber
	\\
	&&\frac12\Big(M_{4}+N_{4}+{M_{4}}
	{\big|}_{\Box_1\leftrightarrow\Box_2}+{N_{4}}
	{\big|}_{\Box_1\leftrightarrow\Box_2}\Big)
	=
  	-\frac1{45\Box_1}-\frac1{45\Box_2}+\frac{\Box_3}
	{45\Box_1\Box_2} , \nonumber
	\\
	&&\frac12\Big(M_{5}+N_{5}+{M_{5}}
	{\big|}_{\Box_1\leftrightarrow\Box_2}+{N_{5}}
	{\big|}_{\Box_1\leftrightarrow\Box_2}\Big)
 	=
   	-\frac1{45\Box_1\Box_2} ,\nonumber
	\\
	&&M_{6}+N_{6} =0 ,\,\,\,\,M_{7}+N_{7}= 0 ,\nonumber
	\\
	 &&\frac12\Big(M_{8}+N_{8}+{M_{8}}
	{\big|}_{\Box_1\leftrightarrow\Box_2}+{N_{8}}
	{\big|}_{\Box_1\leftrightarrow\Box_2}\Big) =
-\frac12,\nonumber
	\\
	&&\frac12\Big(M_{9}+N_{9}+{M_{9}
	{\big|}_{\Box_1\leftrightarrow\Box_2}}
	+{N_{9}}
	{\big|}_{\Box_1\leftrightarrow\Box_2}\Big) =
-\frac1{12},\nonumber
	\\
	&&\frac12\Big(M_{10}+N_{10}+{M_{10}}
	{\big|}_{\Box_1\leftrightarrow\Box_2}+{N_{10}}
	{\big|}_{\Box_1\leftrightarrow\Box_2}\Big) = 0 .
\label{3.28}
	\end{eqnarray}
\arraycolsep=5pt
\mathindent=\parindent
The symmetrizations on the left-hand sides of
these equations correspond to the symmetries of the
tensor structures (\ref{3.23}). With the form
factors (\ref{3.28}), Eq. (\ref{3.27}) takes precisely the form of
the trace
anomaly (\ref{3.11}).

{}From the point of view of the expectation-value equations, one
should also be
able to carry out the calculation of the anomaly in terms of the
spectral
originals. In the derivation above this problem appears only with
respect to
the linear combinations (\ref{3.26}) of the third-order form factors.
Since the
coefficients of these linear combinations contain $\Box$'s, a
calculation in
terms of integral originals encounters difficulties. However, the
generalized
spectral representation based on Eqs.(\ref{2.30})-(\ref{2.31}) makes
the
calculation feasible. The respective technique is worked up in the
report
\onlinecite{IV}. The amount of calculations with the spectral forms
is much
smaller than with the explicit forms of the functions $\Gamma_i$, and
the
result is again Eq.(\ref{3.28}).

In the calculation above, one can trace the two types of anomalies
discussed in
Ref.\onlinecite{DesSch}. Eq. (\ref{3.22}) is the contribution to the
anomaly
stemming from the scale dependent log's in the quadratic terms of the
effective
action, and Eq.(\ref{3.25}) is the contribution coming from varying
the
curvature in the finite cubic terms. Both contributions reduce,
however, to the
{\em scale-independent} log's in Eq. (\ref{3.24}). These nonlocal
log's cancel
{\em only in the sum of the two contributions} in Eq. (\ref{3.27})
leaving the
trees (\ref{3.28}). Trees were the only type of nonlocal form factors
used in
the constructions of the previous works \cite{Riegert,Tomb,Mott}
whereas in
fact we have to do here with functions of three independent arguments
$\Gamma_i(-\Box_1,-\Box_2,-\Box_3)$, and no dimensional
considerations can fix
the dependence of these functions on the ratios $\Box_n/\Box_m$.

\section{Heat kernel and the trace anomaly}
\indent
The trace anomaly can also be derived by making the conformal
transformation in
the heat kernel which contributes to the effective action (\ref{2.4})
via
Eq.(\ref{2.12}).
To enable a comparison with the effective action (\ref{2.6}) ,
one must subtract from the heat kernel (\ref{2.13}) the terms of
zeroth and
first order in the curvature (see a remark to Eq. (\ref{2.6}) in
Ref.\onlinecite{f0}).
For ${\rm Tr}\,K(s)$ with the lowest-order terms subtracted we
introduce the
notation
	\begin{equation}
	{\rm Tr}\, K'(s)={\rm Tr}\, K(s)-
	\frac1{(4\pi s)^2}
	\int\! dx\, g^{1/2}\, {\rm tr}\,(\hat{1}+s\hat{P}).
\label{4.1}
	\end{equation}
The second-order terms in ${\rm Tr}\,K(s)$ transform like in
(\ref{3.15}) but,
instead of $\gamma(-\Box)$, one has to deal with the form factors
	\begin{equation}
	f(-s\Box),\,\,\,\frac{f(-s\Box)-1}{s\Box},\,\,\,\,
	\frac{f(-s\Box)-1-\frac16s\Box}{(s\Box)^2},       \label{4.2}
	\end{equation}
the form factors $f_i(-s\Box)$, $i=1,...5$, in (\ref{2.13}) beeing
the linear
combinations of (\ref{4.2}) with numerical coefficients \cite{II,IV}.
Here
	\begin{equation}
	f(\xi)=\int_{\alpha\geq0}\!d^2\alpha\,
	\delta(1-\alpha_1-\alpha_2)\exp(-\alpha_1\alpha_2\xi)
	=\int^1_0\!d\alpha\,{\rm e}^{-\alpha(1-\alpha)\xi}
\label{4.3}
	\end{equation}
is a fundamental second-order form factor in the trace of the heat
kernel.
The counterparts of Eqs. (\ref{3.18}) and (\ref{3.20}) are in this
case
	\begin{eqnarray}
	&&\int\! dx\, g^{1/2}\,R_{\mu\nu}[f(-s\Box),
	\nabla^\mu\nabla^\nu]\sigma \nonumber
	\\&&\,\,\,\,\,\,\,\,\,\,\,\,\,\,\,\,
	=\int\! dx\, g^{1/2}\,\frac{f(-s\Box_1)-f(-s\Box_3)}
	{\Box_1-\Box_3}[\Box_3,\nabla^\mu_3\nabla^\nu_3]
	R_{1\mu\nu}\sigma_3+{\rm O}[R^3..],
\label{4.4}
	\\
	&&\int\! dx\, g^{1/2}\, {\rm tr}\,
	\Re_1\Big(\delta_\sigma f(-s\Box_2)\Big)\Re_2 \nonumber
	\\&&\,\,\,\,\,\,\,\,\,\,\,\,\,\,\,\,
	=\int\! dx\, g^{1/2}\, {\rm
tr}\,\frac{f(-s\Box_1)-f(-s\Box_2)}
	{\Box_1-\Box_2}(\delta_\sigma\Box_2)\Re_1\Re_2
	+{\rm O}[\Re^3].
\label{4.5}
	\end{eqnarray}
The third-order terms in ${\rm Tr}\,K(s)$ transform in a
way completely similar to the above \cite{IV}. An important
distinction from the previous case is that the linear
nonlocal terms do not cancel.

The total result for ${\rm Tr} K'(s)$ divided by $s$ is
of the form
	\begin{eqnarray}
	\frac1s\delta_\sigma{\rm Tr} K'(s) &=&
	\frac1{(4\pi)^2}\int\! dx\, g^{1/2}\, {\rm tr}\,
	\Big\{\sigma\Box t_1(s,\Box)\hat{P}
	+\sigma\Box t_2(s,\Box)R\hat{1}   \nonumber
	\\
	&&\,\,\,\,\,\,\,\,\,\,\,\,\,\,\,\,\,\,\,\,\,\,\,\,
	+\sum^{10}_{i=1}T_i(s,\Box_1,\Box_2,\Box_3)
	\Re_1\Re_2\sigma_3({i})\Big\}+{\rm O}[\Re^3]    \label{4.6}
	\end{eqnarray}
where $\Re_1\Re_2\sigma_3({i})$ are the tensor structures
(\ref{3.23}), and the functions $t_1, t_2, T_i$ are obtained as
certain
combinations of the form factors in the heat kernel. The differential
equations for these form factors \cite{JMP2} can next be used the
same way as
in two dimensions \cite{JMP2} to bring the functions $t_1, t_2, T_i$
to the form of total derivatives
	\begin{equation}
	t_1=\frac{d}{ds}\widetilde{t}_1,\ \ \ \
	t_2=\frac{d}{ds}\widetilde{t}_2,\ \ \ \
	T_i=\frac{d}{ds}\widetilde{T}_i.
\label{4.7}
	\end{equation}
Here
	\begin{eqnarray}
	&&\widetilde{t}_1 = \frac{f(-s\Box)-1}{s\Box},   \label{4.8}
	\\
	&&\widetilde{t}_2 = \frac1{12}
	\frac{f(-s\Box)-1}{s\Box}-\frac12 \frac{f(-s\Box)
	-1-\frac16s\Box}{(s\Box)^2},
\label{4.9}
	\end{eqnarray}
and the full results for $\widetilde{T}_i$ can be found in the report
\onlinecite{IV}.
Here we need only the asymptotic behaviours of the form factors at
large and
small $s$ of Ref.\onlinecite{JMP2}. Indeed,  the conformal variation
of the
effective action (\ref{2.6}) can now be obtained as
	\begin{equation}
	-\delta_\sigma W = \frac12
	\int^\infty_0\frac{ds}s\delta_\sigma{\rm Tr} K'(s).
\label{4.10}
	\end{equation}
{}From (\ref{4.6}) and (\ref{4.7}) we find
	\begin{eqnarray}
	-\delta_\sigma W &=& -
	{\frac1{2(4\pi)^2}\int\! dx\, g^{1/2}\, \,{\rm tr}\,}
	\Big\{\sigma\Box\widetilde{t}_1(0,\Box)\hat{P}
	+\sigma\Box\widetilde{t}_2(0,\Box)R\hat{1}  \nonumber
	\\
	&&\,\,\,\,\,\,\,\,\,\,\,\,+\sum^{10}_{i=1}\widetilde{T}_i
	(0,\Box_1,\Box_2,\Box_3)
	\Re_1\Re_2\sigma_3({i})\Big\}+{\rm O}[\Re^3],
\label{4.11}
	\end{eqnarray}
where use is made of the asymptotic behaviours at large $s$ to
conclude that
the functions $\widetilde{t}_1,\,\widetilde{t}_2,\,\widetilde{T}_i$
vanish at
$s\rightarrow\infty$. By taking from  Ref.\onlinecite{JMP2} the
asymptotic
behaviours at small $s$, we obtain
	\begin{eqnarray}
	&&\widetilde{t}_1(0,\Box) = \frac16,\,\,\,\,\,
	\widetilde{t}_2(0,\Box) = \frac1{180},  \nonumber
	\\
	&&\frac12\Big(\widetilde{T}_{1}+{\widetilde{T}_{1}}
	{\big|}_{\Box_1\leftrightarrow\Box_2}\Big) = 0,\ \ \ \  s=0,
\nonumber
	\\
	&&\frac12\Big(\widetilde{T}_{2}+{\widetilde{T}_{2}}
	{\big|}_{\Box_1\leftrightarrow\Box_2}\Big) =
\frac1{180}+\frac{\Box_1}
	{180\Box_2}+\frac{\Box_2}{180\Box_1}     \nonumber
	\\
	&&\,\,\,\,\,\,\,\,\,\,\,\,\,\,\,\,\,\,\,\,\,\,
	\,\,\,\,\,\,\,\,\,\,\,\,\,\,\,\,\,\,\,\,\,\,\,\,\,\,
	\,\,\,\,\,\,\,\,\,\,\,\,\,\,\,\,\,\,\,\,\,
	-\frac{\Box_3}{90\Box_1}-\frac{\Box_3}{90\Box_2}
	+\frac{{\Box_3}^2}
	{180\Box_1\Box_2} ,\ \ \ \  s=0,      \nonumber
	\\
	&&\widetilde{T}_{3} =0 ,\ \ \ \
	\frac12(\widetilde{T}_{4}+{\widetilde{T}_{4}}
	{\big|}_{\Box_1\leftrightarrow\Box_2}) =
  	\frac1{45\Box_1}+\frac1{45\Box_2}
	-\frac{\Box_3}{45\Box_1\Box_2} ,\ \ \ \  s=0,  \nonumber
	\\
	&&\frac12\Big(\widetilde{T}_{5}+{\widetilde{T}_{5}}
	{\big|}_{\Box_1\leftrightarrow\Box_2}\Big) =
   	\frac1{45\Box_1\Box_2} ,\ \ \ \
	\widetilde{T}_{6} = 0 ,\ \ \ \
	\widetilde{T}_{7} = 0 ,\ \ \ \  s=0, \nonumber
	\\
	&&\frac12\Big(\widetilde{T}_{8}+{\widetilde{T}_{8}}
	{\big|}_{\Box_1\leftrightarrow\Box_2}\Big) =\frac12,\ \ \ \
\frac12\Big(\widetilde{T}_{9}+{\widetilde{T}_{9}}
	{\big|}_{\Box_1\leftrightarrow\Box_2}\Big)
	=\frac1{12},\ \ \ \ s=0,
\nonumber
	\\
	&&\frac12\Big(\widetilde{T}_{10}+
	{\widetilde{T}_{10}}
	{\big|}_{\Box_1\leftrightarrow\Box_2}\Big) =0,\ \ \ \  s=0.
\label{4.12}
	\end{eqnarray}
With these expressions inserted in (\ref{4.11}), one arrives at
Eq.(\ref{3.11})
which
is the correct trace anomaly.

Because the conformal transformation is inhomogeneous
in the curvature, the expansion in powers of the
curvature does not preserve the exact conformal
properties of the effective action. These
properties can only be recovered order by
order. One can try to remove this shortcoming of
covariant perturbation theory by using the ideas
of Ref. \onlinecite{FV-conf} but such an improvement is
already beyond the scope of the present paper.

\section*{Acknowledgments}

\indent
The work of A.O.B. on this paper was partially supported by the CITA
National Fellowship and NSERC grants at the University of Alberta.
G.A.V. has been supported by the Russian Science Foundation under
Grant 93-02-15594 and by
a NATO travel grant (CRG 920991). V.V.Z., whose work was
supported in part by National Science Council of the Republic of
China under contract No. NSC 82-0208-M008-070, thanks the Department
of Physics
of National Central University for the support in computing
facilities.


\begin{thebibliography}{99}
\bibitem{f0} Since these terms are local and at most quadratic in
	derivatives, they must be removed by renormalization.
	The zeroth-order term violates the boundary condition
	of asymptotic flatness but, in the present case of massless
	quantum fields, it is cancelled by the contribution
	of the functional measure (see Ref.\onlinecite{Bern}).
\bibitem{files} These files are available from the authors.
\bibitem{f1} The dimensional regularization was used in
Ref.\onlinecite{II}
	for the derivation of the quadratic terms in $W$.
	In fact, important is only belonging of the
	regularization to one of the two alternative
	classes discussed in Ref.\onlinecite{FV-conf}.
\end{thebibliography}
\end{document}